\documentclass[10pt]{iopart}

\usepackage{iopams}  
\usepackage{graphicx}  
\usepackage{subfigure}
\usepackage{dcolumn}   
\usepackage{bm}        
\usepackage{amssymb}   
\usepackage{comment}
\usepackage{hyperref}
\usepackage{color}
\usepackage{mathrsfs}
\usepackage{mathcomp}
\usepackage{textcomp}
\usepackage{dsfont}
\usepackage{esint}
\usepackage{braket}
\usepackage{textgreek}
\usepackage{lipsum}

\begin{document}
\newcommand{\red}[1]{\textcolor{blue}{#1}} 
\title[Ultrasonic elastic responses in monopole lattice]{Ultrasonic elastic responses in monopole lattice}

\author{Xiao-Xiao Zhang}
\address{Department of Applied Physics, The University of Tokyo, 7-3-1 Hongo, Bunkyo-ku, Tokyo 113-8656, Japan}
\ead{zhang@appi.t.u-tokyo.ac.jp}
\author{Naoto Nagaosa}
\address{Department of Applied Physics, The University of Tokyo, 7-3-1 Hongo, Bunkyo-ku, Tokyo 113-8656, Japan}
\address{RIKEN Center for Emergent Matter Science (CEMS), 2-1 Hirosawa, Wako, Saitama 351-0198, Japan}
\ead{nagaosa@ap.t.u-tokyo.ac.jp}

\vspace{10pt}
\begin{indented}
\item[]March 2017
\end{indented}

\newcommand\dd{\mathrm{d}}
\newcommand\ii{\mathrm{i}}
\newcommand\ee{\mathrm{e}}
\makeatletter
\def\ExtendSymbol#1#2#3#4#5{\ext@arrow 0099{\arrowfill@#1#2#3}{#4}{#5}}
\newcommand\LongEqual[2][]{\ExtendSymbol{=}{=}{=}{#1}{#2}}
\newcommand\LongArrow[2][]{\ExtendSymbol{-}{-}{\rightarrow}{#1}{#2}}
\newcommand{\cev}[1]{\reflectbox{\ensuremath{\vec{\reflectbox{\ensuremath{#1}}}}}}

\newcommand{\mycomment}[1]{} 

\begin{abstract}
The latest experimental advances have extended the scenario of coupling mechanical degrees of freedom in chiral magnets (MnSi/MnGe) to the topologically nontrivial skyrmion crystal and even monopole lattices. Equipped with a spin-wave theory highlighting the topological features, we devise an interacting model for acoustic phonons and magnons to explain the experimental findings in a monopole lattice with a topological phase transition, i.e., annihilation of monopole-antimonopole pairs. We reproduce the anisotropic magnetoelastic modulations of elastic moduli: drastic ultrasonic softening around the phase transition and a multi-peak-and-trench fine structure for sound waves parallel and orthogonal to the magnetic field, respectively. Comparison with experiments indicates that the magnetoelastic coupling induced by Dzyaloshinskii-Moriya interaction is comparable to that induced by the exchange interaction. Other possibilities such as elastic hardening are also predicted. The study implies that the monopole defects and their motion in MnGe play a crucial role.
\end{abstract}

%
%
%
\maketitle
%
%

\section{Introduction}

Topology is now playing a more and more significant role in condensed matter physics. Apart from an upsurge in the focus on topological classification of quantum phases of matter
, another field bearing ideas of topology, magnetism inhabited by stripes, vortices, domain walls, etc., has been under experimental and theoretical investigations for a long history\cite{bubble}. In the recent decade, the realization of topologically nontrivial Skyrmion or Skyrmion crystal (SkX) in chiral magnets\cite{Skprediction1,Skprediction2,MnSi1,MnSi2,FeGe,FeCoSi1,FeCoSi2,CuOSeO,Fe_film,creation0} revives the old idea originally proposed as a hadron model\cite{Skyrme}, serving as a new scenario of the interplay between the orbital and spin of electrons and ions, and offering plenty of brand new phenomena\cite{THE1,THE2,Zang,SkHall,NFL,Zang2,currentmotion,ultralow1,ultralow2,Zang1} together with the potential for application in magnetic storage\cite{memofunc0,racetrack,memofunc1,memofunc2,memofunc3,memofunc4,memofunc5}. Symmetry breaking of spins in noncentrosymmetric chiral magnets, which bears both the Heisenberg exchange interaction (EXI) and the Dzyaloshinskii-Moriya interaction (DMI)\cite{Dzyaloshinskii,Moriya,Fert&LevyDMI} due to spin-orbit coupling, can embody the Skyrmion texture\cite{Skprediction1,Skprediction2}. 
A minimal Hamiltonian for isolated Skyrmions or a Skyrmion lattice in $d$ spatial dimensions includes the EXI, the Bloch-type DMI, and the Zeeman energy\cite{modelH2,spinwave0} (setting $\hbar=1$ throughout this paper)
\begin{eqnarray}\label{H_SkX}
\mathcal{H}_\mathrm{SkX}= \int{ \dd^d\vec{r}  \left[ \frac{J\mycomment{\hbar^2}}{a_0^{d-2}} \left(\nabla\vec{S}\right)^2  + \frac{D\mycomment{\hbar^2}}{a_0^{d-1}} \vec{S}\cdot\left(\nabla\times\vec{S}\right)  - \frac{\mycomment{\hbar}1}{a_0^{d}} \mu \vec{S}\cdot\vec{B} \right]}
\end{eqnarray}
in which $a_0$ is the microscopic lattice constant. In two-dimensional (2D) thin films, usually magnetic anisotropies can also play a key role in stabilizing the Skyrmion phase.
Dissimilar to other modulated magnetic structures like the conical and helical phases that can be realized therein, a Skyrmion winds a sphere certain times and is characterized by the topological Skyrmion number\cite{Rajaraman}. 
In such magnetic systems, the involvement of topology is often enhanced or even induced by strong correlation effect. Indeed, based on an adiabatic approximation for the real-space Berry phase produced by the fixed-length spin texture of Skyrmion, the constraint drawn by the strong coupling with itinerant electrons can be described by the emergent electromagnetic fields (EEMF)\cite{EEMF0,EEMF1,EEMF2,Zang,Resistivity}.

Apart from the most common Skyrmion lattice as vertical magnetic field induced triangular lattices of Skyrmions observed in chiral magnets, the coalescence or bisection of columnar Skyrmion tubes has been observed experimentally in a bulk $\textrm{Fe}_{1-x}\textrm{Co}_x\textrm{Si}$ material\cite{SkMerge}. Such merging points are in fact singularities or defects (the Bloch points\cite{Bloch2,Bloch1}) in the spin texture 
\begin{equation*}
\vec{n} = \vec{S}/|\vec{S}|
\end{equation*}
where the spin moment $\vec{S}=\vec{0}$. Specifically, they can be regarded as monopoles or antimonopoles\cite{SkMerge,Resistivity,Nii2} in terms of the EEMF and can create or annihilate Skyrmions, which are the tubes in three dimensions (3D). In a Skyrmion lattice with these defects, the Skyrmion tubes may not penetrate the sample and can go to an end at the monopole defects at a depth. For simplicity, we henceforth may use 'monopole' to refer to both monopoles and antimonopoles. A two-Skyrmion-merging model based on the nonlinear sigma model has been used to study its effect on tranport\cite{monopole3}.
Another regular solution for such ended Skyrmions has been investigated in the conical phase\cite{ChiralBlochPoint}. Based on the Landau-Lifshitz-Gilbert equation, people also studied the dynamics and energetics of the monopole defects\cite{monopole1,monopole2}.

Partly owing to the intrinsic 3D nature of the monopole defects, there turns out to be no experimental or theoretical evidence for the formation of a lattice of the monopoles in thin films\cite{Skprediction1,Skprediction2,Bogdanov1,Bogdanov2,Bogdanov3}. 
Nevertheless, it should be more feasible to be realized in 3D bulk chiral magnets. First being theoretically suggested as a superposition of multiple non-coplanar spin spirals\cite{SkX1}, a 3D lattice of monopole defects found corroboration from the thermodynamic calculation followed\cite{SkX2}. Experimentally confirming these expectations, from the analysis of the electric and thermal Hall effects and the Nernst effect in contrast to those of conventional MnSi-like ones, a 3D bulk MnGe under pressure was strongly inferred to possess Skyrmion tubes and a simultaneous lattice of the monopole-antimonopole defects\cite{Kanazawa1,Kanazawa2,Kanazawa3}. Notably, with a lattice variant of the minimal Hamiltonian Eq.~\eref{H_SkX}, latest Monte Carlo simulation confirmed the formation of a monopole-antimonopole lattice in 3D as long as the magnetic period becomes short\cite{Han_monopole_simulation}, which is just the situation of MnGe observed in those experiments. This endorses our inclusion of the EXI and DMI in Sec.~\ref{models} among other possible magnetic interactions. Further, a recent study using Lorentz transmission electron microscopy clearly revealed the magnetic structure of MnGe in concordance with the three-orthogonal-spiral model anticipated\cite{Kanazawa4}. 

It is necessary to pay attention to the distinction between two points of view, i.e., the spin texture $\vec{n}(\vec{r})$ and the original spin moment field $\vec{S}(\vec{r})$, of the magnetic ordering background, upon which we shall mainly study the effect due to fluctuations. The latter, $\vec{S}(\vec{r})$, which is for constructing the multi-spiral spin density waves, is non-singular and topologically trivial because any configuration mapped onto a 3-ball $B^3$ can be smoothly connected to $\vec{S}(\vec{r})=\vec{0}$. However, this work aims at highlighting the effect characteristic for the topologically nontrivial defects and the orientational field $\vec{n}(\vec{r})$, which is inherently described by the second homotopy group of a 2-sphere $S^2$ that cannot shrink to a point, naturally appears to be physically more relevant. On the other hand, for MnGe, in the strong correlation regime, localized spins' moments can hardly vary in magnitude and the saturated magnetization is large and only vanishes at the singular defects. Indeed, this choice of the unit-length field already proved to be appropriate for the description of the influence on conduction electron in the strongly correlated MnGe material, where the 3D monopole lattice even exists in the absence of external magnetic field\cite{Nii2,Resistivity}.

\begin{figure}
\subfigure[$k_\parallel$-mode]{
\label{fig:ExpeIllustxx} 
\begin{minipage}[c]{0.5\textwidth}
\centering
  \scalebox{0.65}{\includegraphics{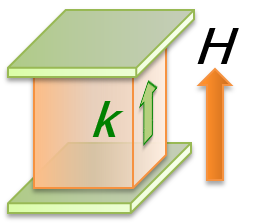}}
\end{minipage}}%
\subfigure[$k_\perp$-mode]{
\label{fig:ExpeIllustzz} 
\begin{minipage}[c]{0.5\textwidth}
\centering
  \scalebox{0.65}{\includegraphics{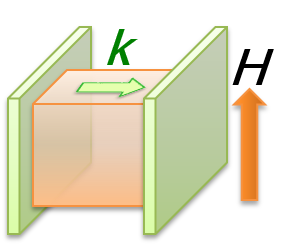}}
\end{minipage}}\\[-10pt]
  \caption{Illustration of experimental settings for the $k_\parallel$-mode and $k_\perp$-mode. $k$ and $H$ stand for the sound wave and the external magnetic field, respectively.}\label{fig:ExpeIllust}
\end{figure}

Expecting new effects due to the monopole defects, we initiated our theoretical investigation to study the phenomena emerging from the coupling between other components and the topologically nontrivial monopole lattice in the hosting material MnGe by calculating the influence from spin-wave excitations therein. Our first study focuses on the longitudinal electric transport in MnGe massively influenced by the quantum fluctuations of the EEMF due to spin waves and identified a topological phase transition of strong correlation genesis\cite{Resistivity}. 
One more intriguing possibility is the coupling to phonons in the hosting solids. Indeed, unconventional ultrasonic responses of elastic stiffness have been observed recently. Experimentalists employed longitudinal sound waves to the Skyrmion phases in both MnSi\cite{Nii1,Petrova} and MnGe\cite{Nii2} with the sound propagating direction parallel ($k_\parallel$-mode) or orthogonal ($k_\perp$-mode) to the applied external magnetic field along the $z$-axis as shown in Fig.~\ref{fig:ExpeIllust}. They recognized the SkX phase by a distinct elastic stiffness with anisotropy for the two modes in MnSi. Dissimilarly, varying with external magnetic field, MnGe showed not only a drastic softening of $k_\parallel$-mode stiffness but also a multi-peak-and-trench fine structure of $k_\perp$-mode stiffness, which are much stronger than those in the SkX phase of MnSi. The range of the magnetic field for the softening considerably coincides with the topological phase transition aforementioned, which is the destruction of the monopole lattice. Fluctuations associated with such qualitative structural transition is therefore not surprising to produce dramatic modification in mechanical properties. 
The EEMF not only helps describe itinerant electrons coupled with localized spins but also 
captures the essential feature of the nontrivial spin texture, hence we find it rather informative even for the magnetoelastic response to be discussed. And it is reasonable to owe the qualitatively different responses of MnGe compared with MnSi to the influence from the monopoles/antimonopoles on the magnon fluctuations therein.

Towards this end, we reuse our spin-wave theory and propose in this paper an interaction theory for magnons and phonons induced by both EXI and DMI, which reproduces and explains the magnetoelastic experimental features. We refer the reader to the separate papers\cite{Resistivity,Nii2} for the detailed description of the monopole lattice consisting of three orthogonal spin spirals, the spin-wave theory of a Skyrmion lattice or monopole lattice, and the experimental confirmation. In Sec.~\ref{models}, we first devise the magnetoelastic interactions and derive the effective phonon theory by integrating out the magnon degrees of freedom. The influence of the magnons on the renormalized phonon excitations is discussed in Sec.~\ref{new_spectrum} with an emphasis on the renormalized phonon linear mode. In Sec.~\ref{Sec:responses} and Sec.~\ref{Sec:spectra}, we compare the experimental observations with our theoretical predictions and explain the physical origin of the magnetic-field-dependent evolution of the hybridized excitation spectra. We emphasize the distinction between the monopole lattice in MnGe and the conventional SkX in MnSi in the rest of Sec.~\ref{result} and conclude in Sec.~\ref{conclusion}.

\section{Theoretical Models}\label{models}
\subsection{Magnetoelastic coupling}\label{ME}
As aforesaid, it is the unit-norm constrained spin texture $\vec{n}=\vec{S}/|\vec{S}|$ that produces the EEMF and yields the nontrivial topology of the Skyrmion and the monopole rather than the bare spin moment $\vec{S}$ itself, which manifests the strong electron correlation therein\cite{Resistivity,Nii2}. If $\vec{S}$ were used, the form factors to be calculated in the following would assume too simple forms to give rise to the complex enough magnetic-field dependence 
and the correct theory to reproduce experiments. We thus make use of $\vec{n}$
instead of $\vec{S}$ to derive the appropriate magnetoelastic interactions from the Hamiltonian Eq.~\eref{H_SkX}. This is the minimal model capable of producing a stable monopole lattice as discussed previously. And this study shows that EXI and DMI are adequate and essential to capture the correct physics herein.

\begin{figure}
\begin{center}
  \scalebox{0.7}{\includegraphics{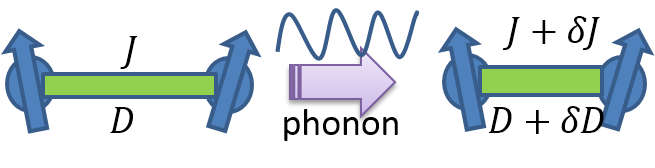}}
\end{center}
\caption{Cartoon for the mechanism of the magnetoelastic coupling.}\label{fig:MEcoupling}
\end{figure}
When longitudinal sound waves are artificially generated in the material, the lattice structure will be perturbed, which is described by the longitudinal phonon. However, the strengths of EXI and DMI depend on lattice bond lengths hence the magnetoelastic coupling (Fig.~\ref{fig:MEcoupling}). Such a perturbation to the magnetic interactions by mechanical degrees of freedom can be accounted for by expanding the strengths of EXI and DMI up to the linear order of phonon degrees of freedom
\numparts
\begin{eqnarray}\label{eq-MEexpansion1}
J (\nabla\vec{n})^2 \rightarrow (J_0+\alpha_{\mathrm{EXI}}(\partial_j u_j)) \partial_j n_i \partial_j n_i \\
\label{eq-MEexpansion2}
D \vec{n}\cdot(\nabla\times\vec{n}) \rightarrow \epsilon^{ijk}\left(D_0+\alpha_{\mathrm{DMI}}(\partial_j u_j)\right) n_i \partial_j n_k
\end{eqnarray}
\endnumparts
wherein $\vec{u}$ is the lattice displacement from equilibrium in the continuum limit, $\alpha_\mathrm{EXI/DMI}$ is the coefficient of this expansion, and summation over repeated indices is understood henceforth. We emphasize four features of this expansion. 1) In conformity with translational symmetry, either $J$ or $D$ depends on $\partial_j u_j$ rather than lattice displacement $\vec{u}$ itself. 2) Alongside, $\partial_j u_j$ obviously deals with the longitudinal phonon. 3) All spatial derivatives match in the direction since any $\partial_j \vec{n}$, reflecting magnetic interactions between two adjacent spins sitting at the endpoints of a lattice bond along the $j$-direction in a lattice model, should be affected by the longitudinal phonon propagating along the same direction through $\partial_j u_j$. 4) As stated in 3), these magnetoelastic couplings take anisotropy into account so as to produce the delicate experimental observations.

The next step is to expand $\vec{n}$ as $n_i = n_i^{(0)} + \varphi_\mu \partial_{\varphi_\mu}n_i^{(0)}$ in Eqs.~\eref{eq-MEexpansion1}\eref{eq-MEexpansion2} to incorporate the spin-wave degrees of freedom $\varphi_\mu = (\vec{\phi}\,,\vec{\delta m})\,,\mu=1,\cdots ,6$ and the superscript $(0)$ means the ground state value or setting $\varphi_\mu=0$ after taking the derivative. For a spin spiral indexed by $i$, this $\phi_i$ is the phase of the constituent spin density wave, which also indicates the spatial shift of the Skyrmion lattice or the position of the monopole defects. And $m_i$ is the magnetization along the propagation axis of spin spiral $i$, generating rotation around this axis. Note that we consider the situation where there are three orthogonal spin spirals and it is those quantities' deviations away from their static mean field values that constitute the spin-wave fields $\varphi_\mu$ in the expansion\cite{Resistivity}. Without loss of generality, we thus set the static value of any $\phi_i$ to $0$ and
denote the fluctuation in $m_i$ by $\delta m_i$.

Starting from the EXI and DMI induced magnetoelastic interaction parts in Eqs.~\eref{eq-MEexpansion1}\eref{eq-MEexpansion2}, up to terms bilinear in spin-wave and phonon fields, one arrives at 
\numparts
\begin{eqnarray}\label{eq-MEDMIEXI1}
\label{eq-MEEXI1}
\textrm{EXI part: } \left[ \partial_{\varphi_\mu}(\partial_jn_i^{(0)}\partial^jn^{i(0)}) \varphi_\mu + 2\partial_{\varphi_\mu} n_i^{(0)} \partial^j n^{i(0)} \partial^j\varphi_\mu \right] \partial_ju_j \\
\label{eq-MEDMI1}
\textrm{DMI part: } \varepsilon^{ijk}\left[ \partial_{\varphi_\mu}(n_i^{(0)}\partial_jn_k^{(0)}) \varphi_\mu + n_i^{(0)}\partial_{\varphi_\mu}n_k^{(0)} \partial_j\varphi_\mu \right] \partial_ju_j,
\end{eqnarray}
\endnumparts
wherein we temporarily omit all the prefactors for simplicity. One crucial criterion is the translational invariance in the continuum model to be derived, i.e., for the spin-wave field $\phi_i$, only its spatial derivative can enter simply because $\phi_i$ is the (phase) displacement field of spin spiral $i$. 
Therefore, when $\mu=1,2,3$, i.e., for the $\phi_i$ field, $\partial_{\varphi_\mu}$ applied upon the parenthesis in the first term in 
either Eq.~\eref{eq-MEEXI1} or Eq.~\eref{eq-MEDMI1} is equivalent to a spatial derivative. And we notice that the typical wave vector of the periodic function in the parenthesis is much larger than that of the slowly varying spin-wave field $\phi_i$. Since Eqs.~\eref{eq-MEEXI1}\eref{eq-MEDMI1} will be part of a Lagrangian density, a spatial integration can be carried assuming $\phi_i$ is uniform, resulting in zero in the end for the total derivative. This procedure exactly eliminates the term linear in $\phi_i$ field, otherwise the translational invariance would be violated. However, this is unnecessary for $\mu=4,5,6$, i.e., for the fluctuating field $\delta m_i$. Therefore, in either Eq.~\eref{eq-MEEXI1} or Eq.~\eref{eq-MEDMI1}, for $\mu=1,2,3$, only the second term remains.

Thus for either EXI or DMI case, we attain two terms
\begin{equation}\label{eq-ME}
C^{\mu j} (\partial_j\varphi_\mu \partial_ju_j)
+ D^{\mu j} (\varphi_\mu \partial_ju_j)
\end{equation} 
 in total, wherein $\mu=1,\dots,6$ but $D^{\mu j} \equiv 0$ whenever $\mu<4$.  
 We call the functions preceding the bilinear fields form factors, which render the coupling between magnon and phonon nonuniform in space. They are defined as $C_\mathrm{EXI}^{\mu j} = 2\partial_{\varphi_\mu} n_i^{(0)} \partial^j n^{i(0)} \,, D_\mathrm{EXI}^{\mu j} = \partial_{\varphi_\mu}(\partial_jn_i^{(0)}\partial^jn^{i(0)}) \,, C_\mathrm{DMI}^{\mu j} = \varepsilon^{ijk} n_i^{(0)}\partial_{\varphi_\mu}n_k^{(0)}\,,D_\mathrm{DMI}^{\mu j} = \varepsilon^{ijk} \partial_{\varphi_\mu}(n_i^{(0)}\partial_jn_k^{(0)})$ and actually record the information of the complicated magnetic structure affecting the EXI or DMI induced magnetoelastic interactions. After Fourier transform to the momentum space, we take the spatial average over a magnetic unit cell of the form factors, making the couplings dependent only on the variable uniform magnetization $m_z$ along the $z$-axis. Therefore we have an exactly solvable theory without couplings between unequal momenta. This uniform magnetization $m_z$ directly relates to the external magnetic field applied on the system along the $z$ axis. The magnetic field dependence of the ultrasonic responses is due to the fact that the magnetic texture of the monopole lattice varies with respect to $m_z$\cite{Resistivity}.
 
\subsection{Effective theory of phonon}\label{ph_eff}
A standard free theory of the longitudinal phonon reads
\begin{equation}\label{eq-ph}
\mathcal{L}_\mathrm{ph} = \frac{1}{2} \left[ c(\partial_\tau \vec{u})^2 + \kappa (\nabla \cdot \vec{u})^2 \right],
\end{equation}
wherein $\kappa$ is an elastic constant, $c$ is the mass density and we work in imaginary time henceforth. For MnGe, spin-wave $\mathcal{L}_\mathrm{SW}$ takes the form\cite{Resistivity}
\begin{equation}\label{eq_Lspinwave}
\mathcal{L}_\mathrm{SW} = \sum_{\alpha=x,y,z} \left[
 \ii \epsilon^{\alpha\beta\gamma}  A b_\alpha \phi_\beta\dot{\phi}_\gamma -\ii B \delta m_\alpha\dot{\phi}_\alpha 
 + \chi{\delta m_\alpha}^2 + \rho (\nabla\phi_\alpha)^2 
 \vphantom{A b_\alpha \phi_\beta\dot{\phi}_\gamma}
 \right]
\end{equation} 
wherein $A = -2q_\mathrm{e} S\frac{1}{k_j k_k} \frac{1}{a_0^{d}}\,, B = \frac{1\mycomment{\hbar}}{a_0^{d}}\,, \chi = \frac{D^2}{J a_0^{d}} \mycomment{\frac{\hbar^2}{a_0^{d}}}\,, \rho = \frac{J}{a_0^{d-2}}\mycomment{J\frac{\hbar^2}{a_0^{d-2}}}$. Note that the $m_z$-dependent $\vec{b}$ is the emergent magnetic field that characterizes the nontrivial topology of the Skyrmions or monopoles in the system. And we get the magnetoelastic interactions
\begin{equation}\label{eq_Lme}
\fl\mathcal{L}_\mathrm{ME} = \frac{\alpha_{\mathrm{EXI}} \mycomment{\hbar^2}}{a_0^{d-2}} (C_\mathrm{EXI}^{\mu j} \partial_j\varphi_\mu \partial_ju_j
+ D_\mathrm{EXI}^{\nu j} \varphi_\nu \partial_ju_j) + \frac{\alpha_{\mathrm{DMI}} \mycomment{\hbar^2}}{a_0^{d-1}} (C_\mathrm{DMI}^{\mu j} \partial_j\varphi_\mu \partial_ju_j
+ D_\mathrm{DMI}^{\nu j} \varphi_\nu \partial_ju_j)
\end{equation} 
 from Sec.~\ref{ME}.
Because the three parts, Eqs.~\eref{eq-ph}\eref{eq_Lspinwave}\eref{eq_Lme}, comprising the full theory are all bilinaear, we can diagonalize the actions in the energy-momentum space. Below, $k$ is used as a shorthand for the argument $(\vec{k},z)$, wherein $z$ is a generic complex frequency that can equal $\ii\omega_n$ for instance. The longitudinal phonon action is transformed to
\begin{equation*}
\mathcal{S}_\mathrm{ph} = \int_0^\beta {\dd\tau  \int {\dd^d \vec{r}  \mathcal{L}_\mathrm{ph}  }  }
= \frac{1}{2} \sum_{k} {u^\mathrm{T}(k) M_\mathrm{ph}(k) u(-k)},
\end{equation*}
in which $3\times3$ matrix $(M_\mathrm{ph})_{ij} = \kappa k_ik_j - cz^2 \delta_{ij}$ and $u = (u_x,u_y,u_z)^\mathrm{T}$.
The spin-wave action is transformed to
\[
\mathcal{S}_\mathrm{SW} = \int_0^\beta {\dd\tau  \int {\dd^d \vec{r}  \mathcal{L}_\mathrm{SW}  }  }
= \frac{1}{2} \sum_{k} {\varphi^\mathrm{T}(k) M_\mathrm{SW}(k) \varphi(-k)},
\]
in which $6\times6$ matrix $M_\mathrm{SW}$ can be solved and $\varphi = (\varphi_1,\dots,\varphi_6)^\mathrm{T}$. The action of magnetoelastic interactions becomes
\begin{equation*}
\mathcal{S}_\mathrm{ME} = \int_0^\beta {\dd\tau  \int {\dd^d \vec{r}  \mathcal{L}_\mathrm{ME}  }  }
= \sum_{k} {F^\mathrm{T}(k) \varphi(-k)},
\end{equation*}
wherein $6$-vector $F(k) = U(k)u(k) = \left(  U_\mathrm{EXI}(k) +  U_\mathrm{DMI}(k) \right) u(k)$, $6\times3$ matrices $(U_\mathrm{EXI})^{\mu j} = \frac{\alpha_\mathrm{EXI} \mycomment{\hbar^2}}{a_0^{d-2}} [(k^j)^2C_\mathrm{EXI}^{\mu j} + \ii k^jD_\mathrm{EXI}^{\mu j}]\,, (U_\mathrm{DMI})^{\mu j} = \frac{\alpha_\mathrm{DMI} \mycomment{\hbar^2}}{a_0^{d-1}} [(k^j)^2C_\mathrm{DMI}^{\mu j} + \ii k^jD_\mathrm{DMI}^{\mu j}] $.

Now the full theory is constituted of three parts $\mathcal{S} = \mathcal{S}_\mathrm{ph}+\mathcal{S}_\mathrm{SW}+\mathcal{S}_\mathrm{ME}$, which are the free phonon theory, the free magnon theory, and their coupling due to the magnetoelastic interactions in turn. 
Next, we can integrate out the harmonic magnon bath in the path-integral formalism, leading to the partition function expressed as
\begin{equation*}
\mathcal{Z}[\varphi,u] = \int{ \mathscr{D}\varphi\mathscr{D}u \: \ee^{-\mathcal{S}\mycomment{/\hbar}} } = \mathcal{Z}[\varphi, u \equiv 0] \int{ \mathscr{D}u \: \ee^{-\mathcal{S}_\mathrm{eff}\mycomment{/\hbar}} }.
\end{equation*}
Because of the bilinearity, after a Gaussian functional integration, we get the effective action for the phonons
\begin{eqnarray}\label{eq-eff}
\mathcal{S}_\mathrm{eff} \mycomment{= \sum_k  \left[ -\frac{1}{2} D^\mathrm{T}(k)M_\mathrm{SW}^{-1}(k)D(-k) + \frac{1}{2} u^\mathrm{T}(k)M_\mathrm{ph}(k)u(-k) \right] \nonumber\\} 
= \sum_k  u^\mathrm{T}(k) \left[ -\frac{1}{2} U^\mathrm{T}(k) M_\mathrm{SW}^{-1}(k) U(-k) + \frac{1}{2} M_\mathrm{ph}(k) \right] u(-k) .
\end{eqnarray}
The expression in the bracket is nothing but $-1$ multiplying the inverse of the $3\times3$ Matsubara Green's function matrix $\mathcal{G}_\mathrm{eff}(k)$ of the effective phonon theory in the energy-momentum space.

\section{Methods and results}\label{result}

\subsection{Renormalized phonon spectrum}\label{new_spectrum}
Instead of the hardly accessible analytic dispersion relations, we resort to inspecting the spectral function $A(\vec{k},\omega)$ of this effective theory. In conformity with experimental investigations, we focus on the effects due to sound waves propagating along $x$ and $z$ directions by 
looking at the diagonal $A_{ii}(k_i,\omega) = -2 \Im G_{ii}^\mathrm{R}$ as a function of momentum $k_i$ along $i$-direction and frequency $\omega$, wherein retarded Green's function $G_{ii}^\mathrm{R}$ comes from the analytic continuation $\left(\mathcal{G}_\mathrm{eff}\right)_{ii}(k_i,z\rightarrow \omega+\ii\eta)$. First of all, we confirm the property of the bosonic spectral function\cite{Mahan} that it is always positive for $\omega>0$ and negative for $\omega<0$. Secondly, the renormalized phonon spectrum is reflected in $A_{ii}(k_i,\omega)$ plot by $\delta$-function-type ridge structures. Thirdly, setting a realistic but small enough $\eta$, we can extract the information of the phonon excitations from the $A_{ii}(k_i,\omega)$ plot by identifying the \mycomment{mildly smoothed }ridge structures.

\begin{figure}
\begin{center}
  \scalebox{0.35}{\includegraphics{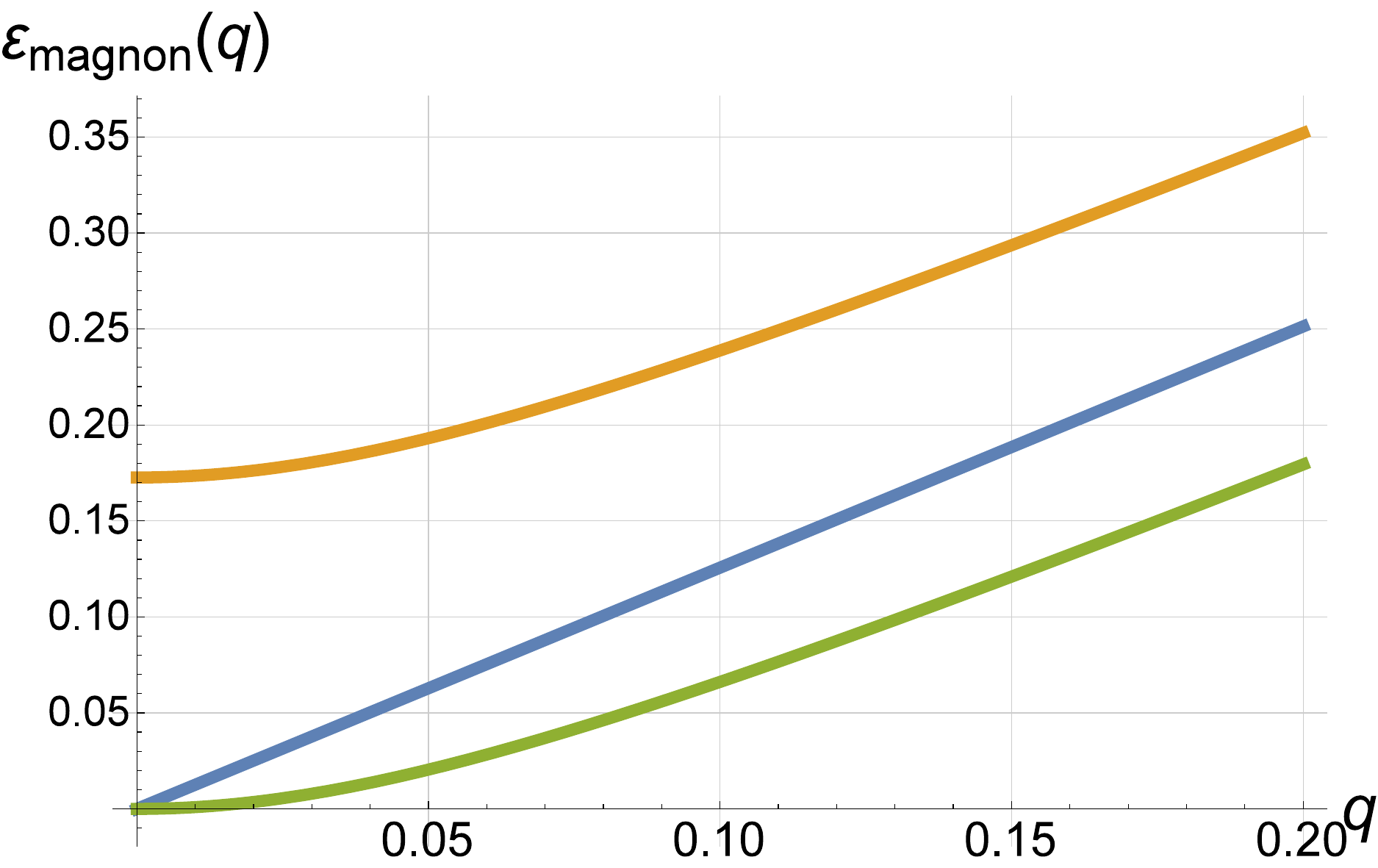}}
  \caption{Low-energy magnon spectrum.}\label{fig_magnon}
\end{center}
\end{figure}

Equation~\eref{eq-ph} itself can only give banal longitudinal phonon excitations of linear dispersion $\omega = v_0 k$ with the acoustic velocity $v_0 = \sqrt{\kappa/c}$. On the other hand, the magnon theory possesses an excitation spectrum composed of three distinct modes\cite{Resistivity}. For MnGe, there are two kinds of gapless excitation, one acoustic mode $\omega_1 = 2D\mycomment{\hbar} a_0 k \propto Dk$ and one quadratic mode $\omega_2 = \frac{\rho}{Ab}k^2 \propto Jk^2$ when $k$ is small, and another excitation with an energy gap $\Delta_\mathrm{mag} = \frac{4\chi Ab}{B^2} \propto \frac{D^2}{J}$, as shown in Fig.~\ref{fig_magnon}. Because of the spin Berry phase of the topologically nontrivial spin textures of Skyrmions/monopoles, any two unparallel spin spirals are possible to mingle with each other. This gives rise to $\phi$-quadratic couplings like $\phi_x\dot{\phi}_y$ in Eq.~\eref{eq_Lspinwave}, mixing up conjugate $\phi$ fields and hence the quadratic mode\cite{Resistivity,Zang,spinwave1}.
An exception is that these three modes degenerate into the linear gapless mode $\omega_1$ when $m_z=0$. These gapless excitations are the Nambu-Goldstone bosons due to the broken translation symmetry. In the case of electric transport\cite{Resistivity}, the softest one affects the electrons' motion the most. However, all the gapless modes will play a crucial role in the low-energy physics describing the interplay with the linear acoustic phonons.

\begin{figure}
\begin{center}
  \scalebox{0.35}{\includegraphics{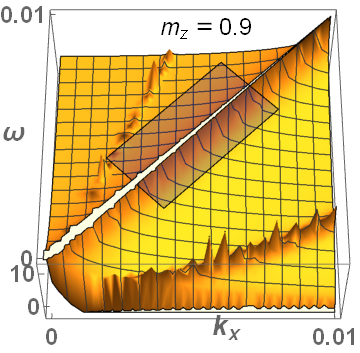}}
  \caption{Scanning the shaded region to extract the renormalized phonon velocity from a typical logarithmic plot of the spectral function $A_{xx}(k_x,\omega)$.}\label{fig:demarcate}
\end{center}
\end{figure}

The salient point is that because of the magnetoelastic coupling, the three magnon modes will hybridize with the phonon mode, giving rise to rich possibilities of renormalized excitation spectra. One interesting instance occurs when the phonon mode intersects with the magnon modes in the dispersion plot, resulting in mutual repulsion and reconstruction of the involved dispersion curves. In the gross, despite possible reconstructions, one is in general still able to relate the new modes to their respective precedents before hybridization, which will henceforth be used as convenient tags of the new modes in the effective phonon theory. Even without any intersection in the dispersion curves, the intensity of a new magnon mode (sharpness of the ridge structure) in a spectral function plot directly reflects the degree of hybridization that influences the (approximately) linear renormalized phonon mode. We have the relation $V_\mathrm{renorm}(m_z) = \sqrt{\kappa_\mathrm{renorm}(m_z)/c}$ (seen from Eq.~\eref{eq-ph}) between the velocity of the renormalized phonon linear mode and the new stiffness $\kappa_\mathrm{renorm}$. This phonon linear mode after hybridization is of the major importance because it is this stiffness $\kappa_\mathrm{renorm}$ that is measured experimentally at different external magnetic fields as the ultrasonic responses. Its velocity is the slope of the corresponding dispersion ridge extracted from scanning the spectral function, e.g., in the shaded region in Fig.~\ref{fig:demarcate}.

Because of the experimental difficulty in determining the accurate value of $J$, $D$ and the magnetoelastic couplings $\alpha_\mathrm{EXI/DMI}$, one has to search for the correct control parameters corresponding to the real materials. 
 We consider two, $\alpha_{\frac{\mathrm{DMI}}{\mathrm{EXI}}}$ and $R_\mathrm{velo}$, for the effective theory. One is the ratio between strengths of different magnetoelastic couplings $ \frac{\alpha_\mathrm{DMI}}{\alpha_\mathrm{EXI}}$, denoted as $\alpha_{\frac{\mathrm{DMI}}{\mathrm{EXI}}}$. The other is the ratio of the velocity of the magnon linear mode, $v_\mathrm{mag}=2D a_0$, to the original unperturbed acoustic velocity $v_0$, denoted as $R_\mathrm{velo} = \frac{v_\mathrm{mag}}{v_0} = \frac{2D a_0}{\sqrt{\kappa/c}}$. We also give an estimate of the characteristic energy scales in the MnGe experiment. The sound wave propagating in MnGe crystal has frequency $f_0 = 18\mathrm{MHz}$ and wavelength $\lambda = 260\mathrm{\mu m}$\cite{Nii2}. The original acoustic velocity is hence $v_0 = \lambda f_0 = 4.7\times10^3 \mathrm{m/s}$ and the phonon energy is $\epsilon_0 = \hbar\omega_0 = 1.2\times10^{-26}\mathrm{J}$. As for the material MnGe\cite{Kanazawa2,Kanazawa4}, $a_0$ is about 4\AA\ and the melting temperature ($\sim200\mathrm{K}$) of the magnetic order can be used to estimate the strength of EXI $J$. We set $J=10D$ for MnGe. Thus, it is straightforward to obtain $v_\mathrm{mag} = 12.0\times10^3\mathrm{m/s}$ and $\Delta_\mathrm{mag} = 2.4\times10^{-23}\mathrm{J}$, using the typical value when $m_z=0.8$. On the other hand, in our theoretical study, we set $\eta = 1\times10^{-6}$, $a_\mathrm{SkX}=2\pi$, $\hbar=\kappa=c=1$ and hence $v_0=1$. Consequently, $J$ or $D$ is fixed by $R_\mathrm{velo}$. In addition to the foregoing variable $\alpha_{\frac{\mathrm{DMI}}{\mathrm{EXI}}}$, we set $\alpha_\mathrm{EXI} = 0.8$ since a too large magnetoelastic coupling unrealistically alters the phonon spectrum while a tiny one renders the effect feeble to detect.

The first message from the above is that $v_\mathrm{mag}$ 
is of the same order as $v_0$. This implies that we had better tune $R_\mathrm{velo}$ not far from unity if we were to explain the experiment. The second message is $\epsilon_0 \ll \Delta_\mathrm{mag}$, which means the low-energy phonon excitations and hence gapless magnons around the long wavelength limit play a major role. 
Because the $U$ matrix in Eq.~\eref{eq-eff} contains two individual parts due to EXI and DMI, 
the combination becomes not simply a summation of the separate effects, but one with inevitable interference between the two types of magnetoelastic interactions. Indeed, we have seen distinct $V_\mathrm{renorm}(m_z)$ profiles when changing $\alpha_{\frac{\mathrm{DMI}}{\mathrm{EXI}}}$ within a typical range $[-5,5]$ by step $0.1$. Note that the sign difference between two magnetoelastic interactions is possible as implied by the sign change in DMI\cite{anticrossing1,anticrossing2,size_control}.

\subsection{Rich possiblities of magnetoelastic responses}\label{Sec:responses}
\begin{figure}
\subfigure[$k_\perp$-mode]{
\label{fig:softenxx} 
\begin{minipage}[c]{0.45\textwidth}
\centering
  \scalebox{0.22}{\includegraphics{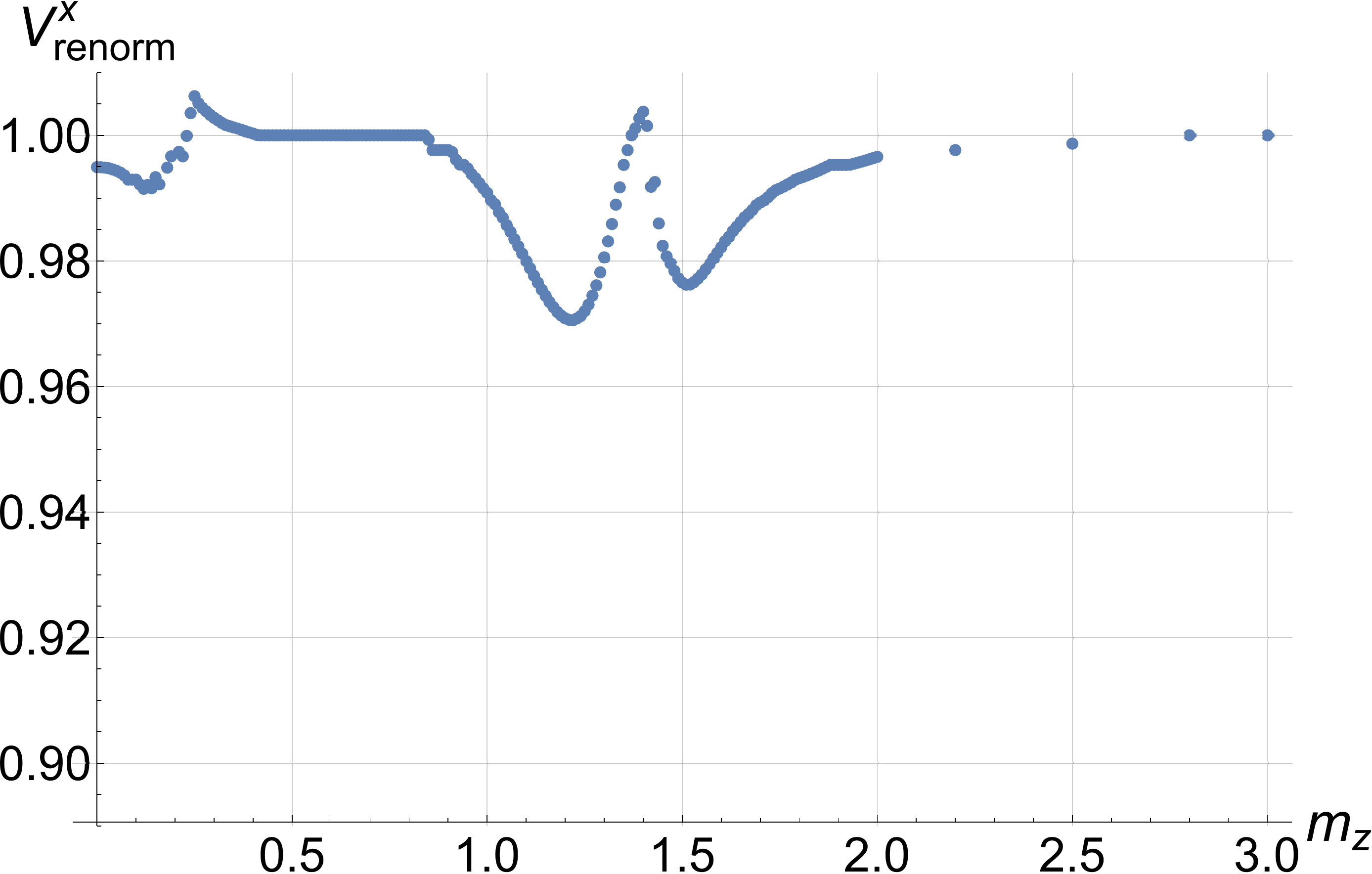}}
\end{minipage}}%
\hfill
\subfigure[$k_\parallel$-mode]{
\label{fig:softenzz} 
\begin{minipage}[c]{0.45\textwidth}
\centering
  \scalebox{0.22}{\includegraphics{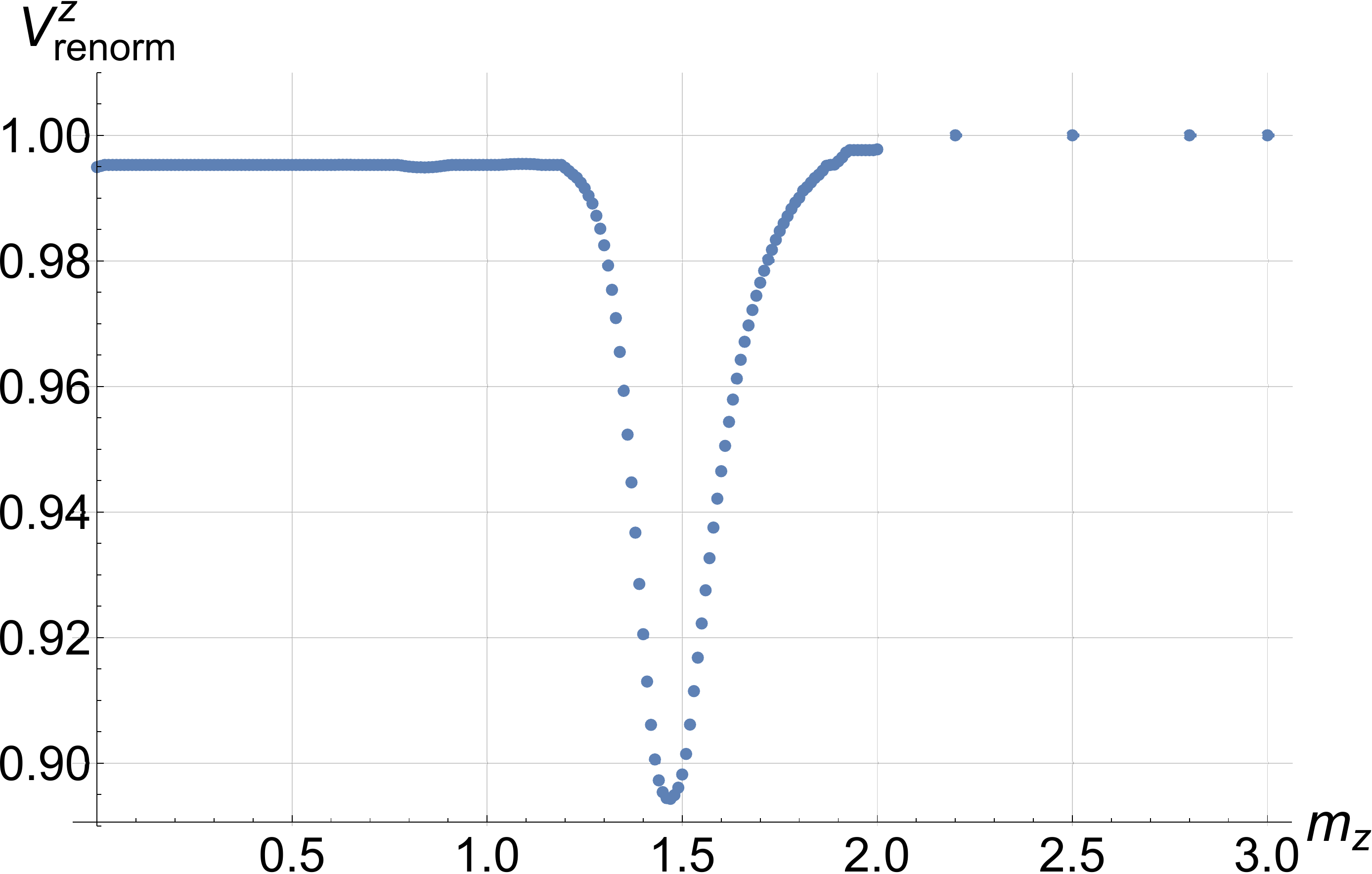}}
\end{minipage}}\\[-10pt]
  \caption{The elastic softening. The velocities of the renormalized linear phonon excitations v.s. the uniform magnetization $m_z$ under control parameters $\alpha_\frac{\mathrm{DMI}}{\mathrm{EXI}} = 1\,,R_\mathrm{velo} = 2.4$. Important experimental features, strong anisotropy, substantial softening, and multi-peak-and-trench fine structure, are theoretically reproduced.}\label{fig:soften}
\end{figure}

\begin{figure}
\subfigure[$k_\perp$-mode]{
\label{fig:softenxxExpe} 
\begin{minipage}[c]{0.45\textwidth}
\centering
  \scalebox{0.34}{\includegraphics{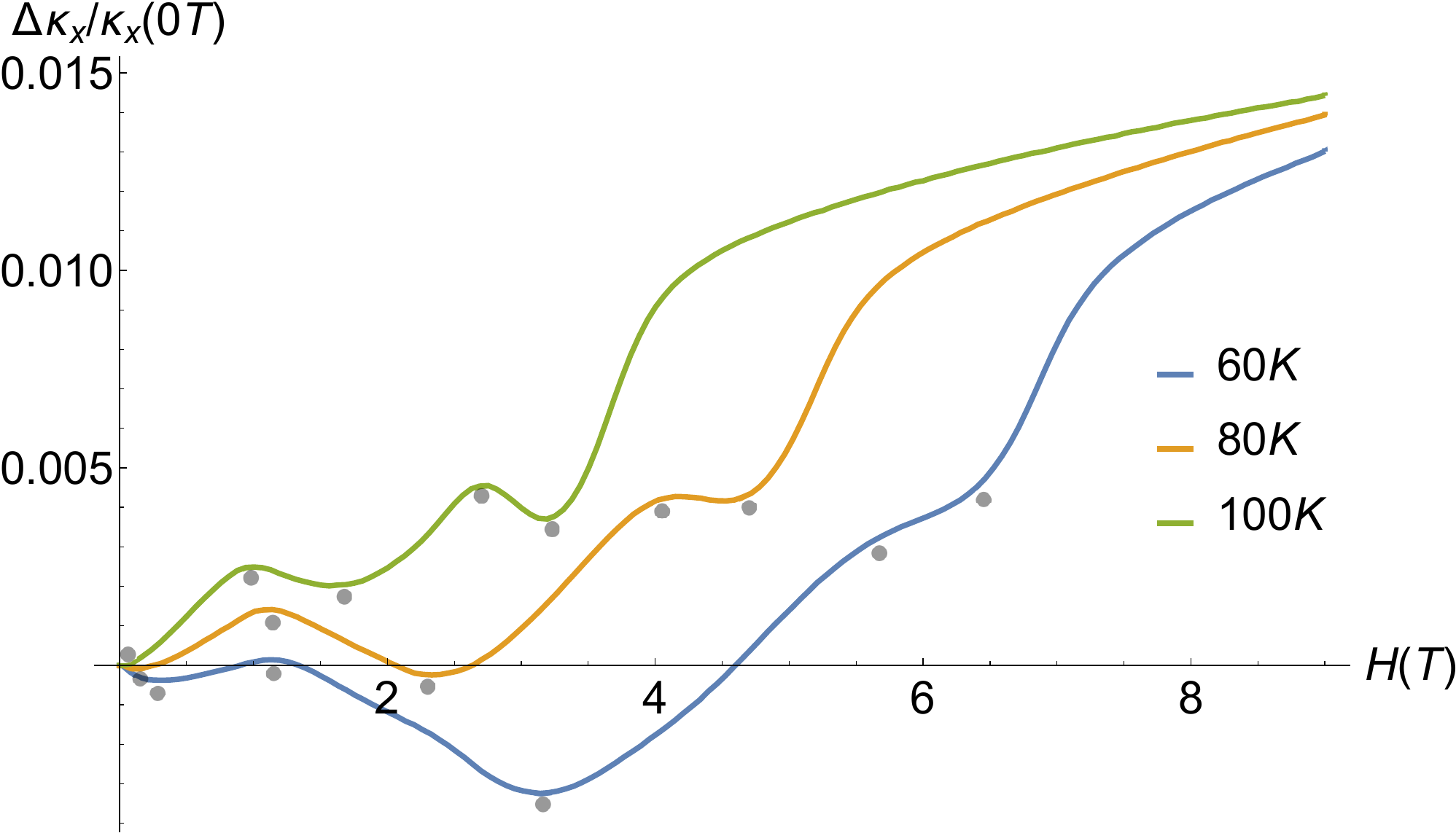}}
\end{minipage}}%
\hfill
\subfigure[$k_\parallel$-mode]{
\label{fig:softenzzExpe} 
\begin{minipage}[c]{0.45\textwidth}
\centering
  \scalebox{0.34}{\includegraphics{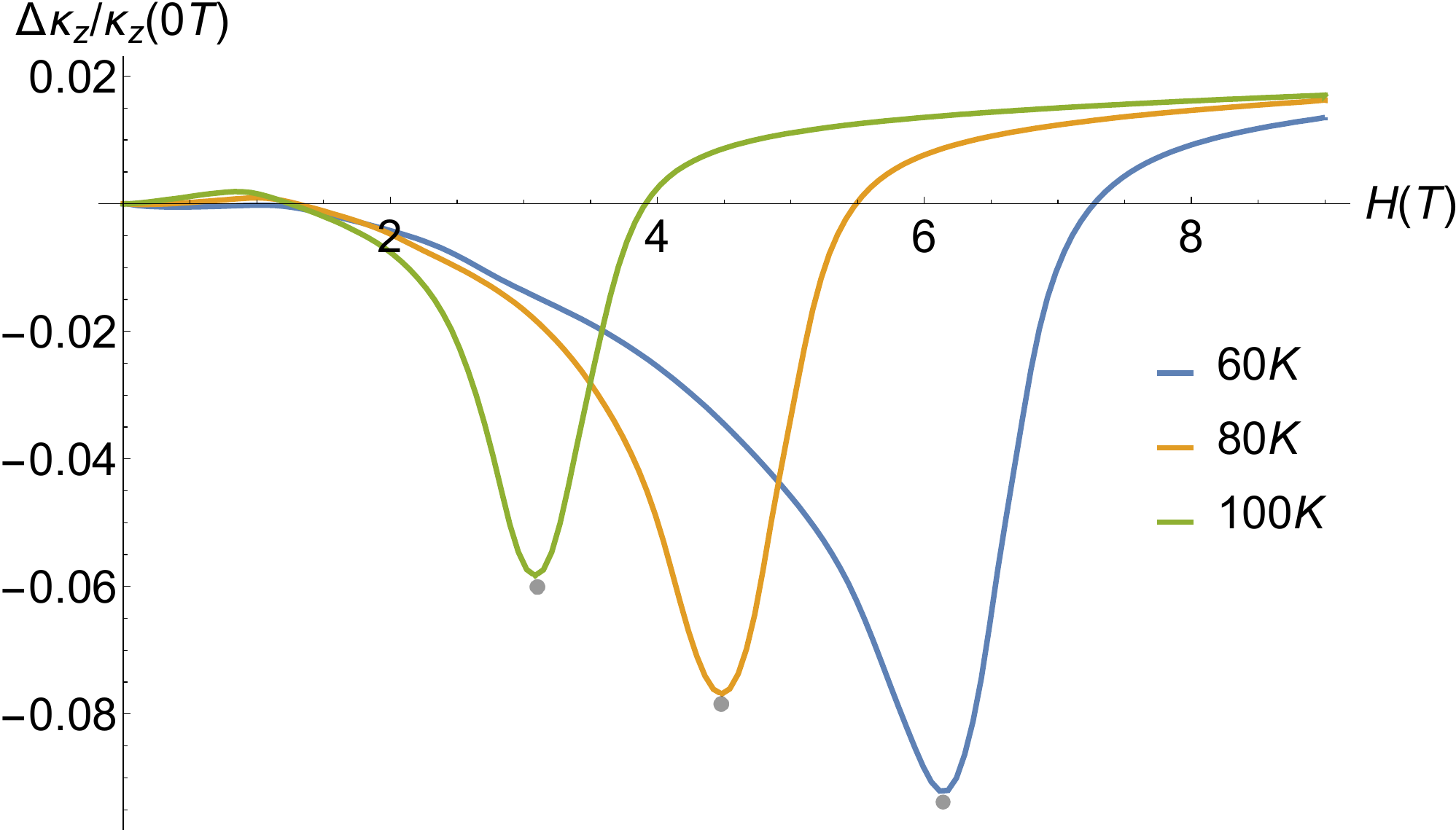}}
\end{minipage}}\\[-10pt]
  \caption{The experimentally observed softening signal at several temperatures. The relative change of the elastic constant $\kappa_{x/z}$ v.s. the external magnetic field $H$ along $z$ direction. Characteristic trench and peak positions are marked by gray dots. Figure plotted from experimental data which are partly reported in the separate paper\cite{Nii2}.}\label{fig:softenExpe}
\end{figure}
\begin{figure}
\subfigure[$\alpha_\frac{\mathrm{DMI}}{\mathrm{EXI}} = -1\,,R_\mathrm{velo} = 0.2$]{
\label{fig:harden1} 
\begin{minipage}[c]{1.0\textwidth}
\centering
  \scalebox{0.155}{\includegraphics{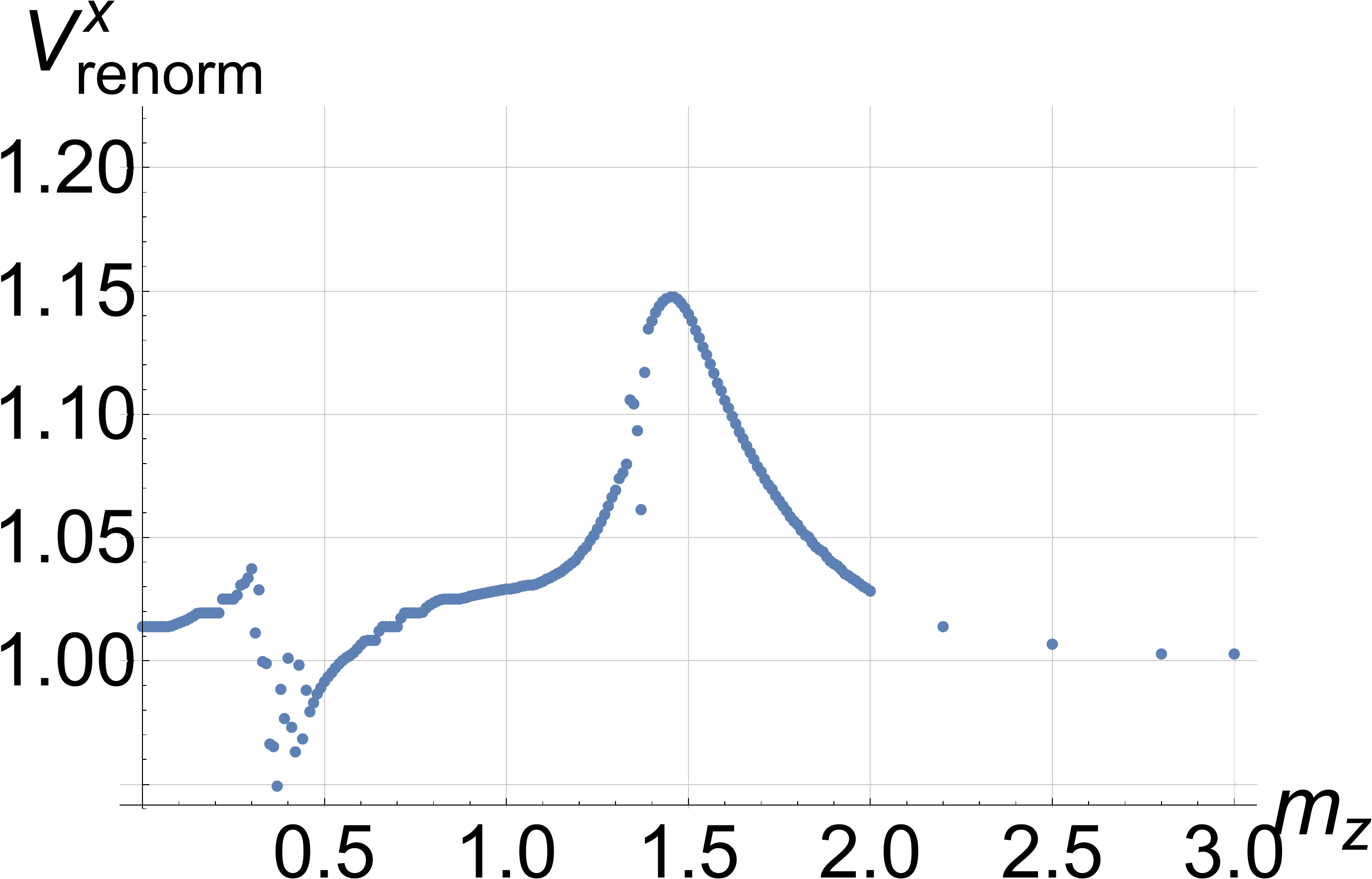}}
  \scalebox{0.155}{\includegraphics{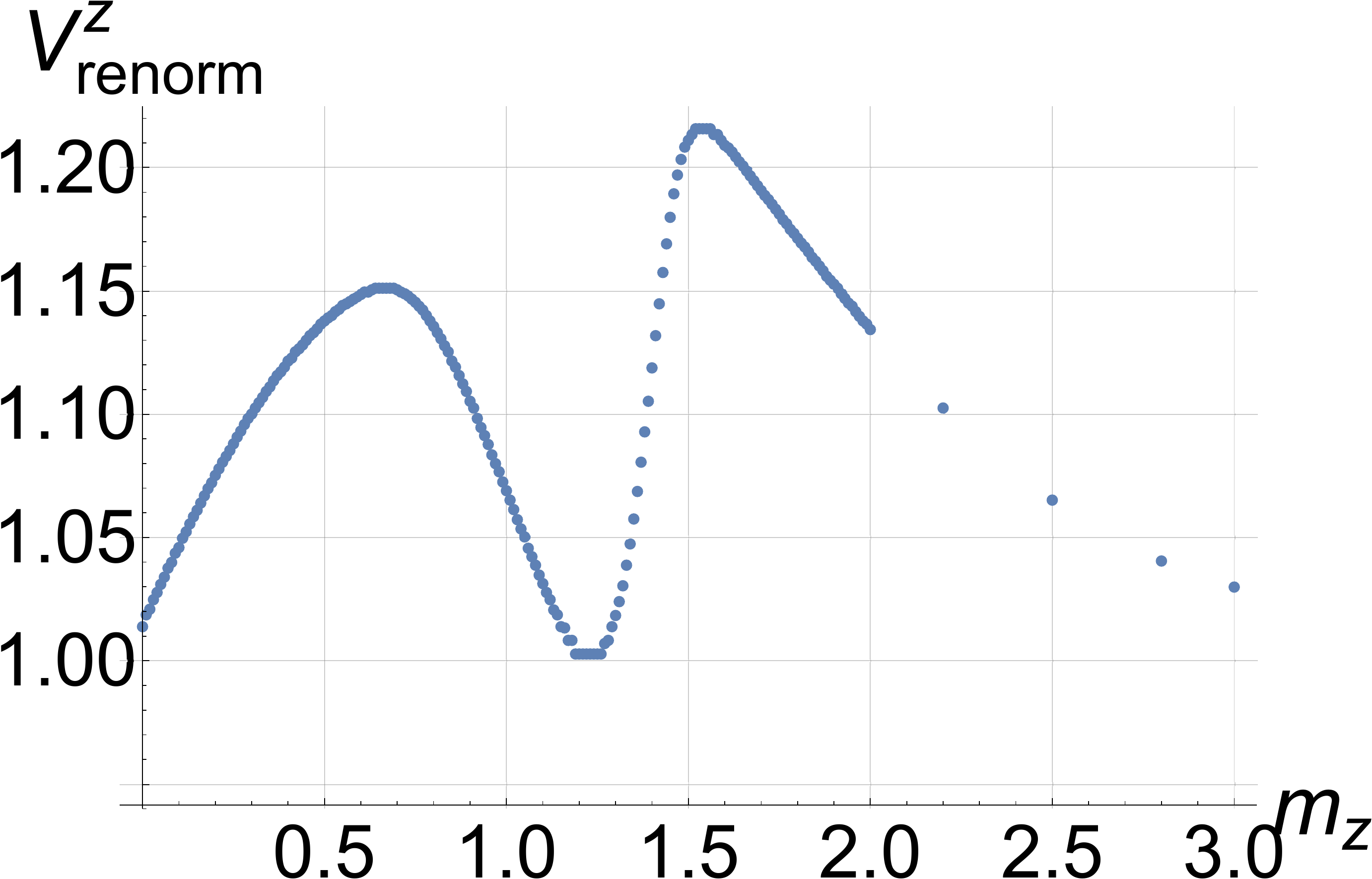}}
\end{minipage}}\\[-7pt]
\subfigure[$\alpha_\frac{\mathrm{DMI}}{\mathrm{EXI}} = 2\,,R_\mathrm{velo} = 0.6$]{
\label{fig:harden2} 
\begin{minipage}[c]{1.0\textwidth}
\centering
  \scalebox{0.155}{\includegraphics{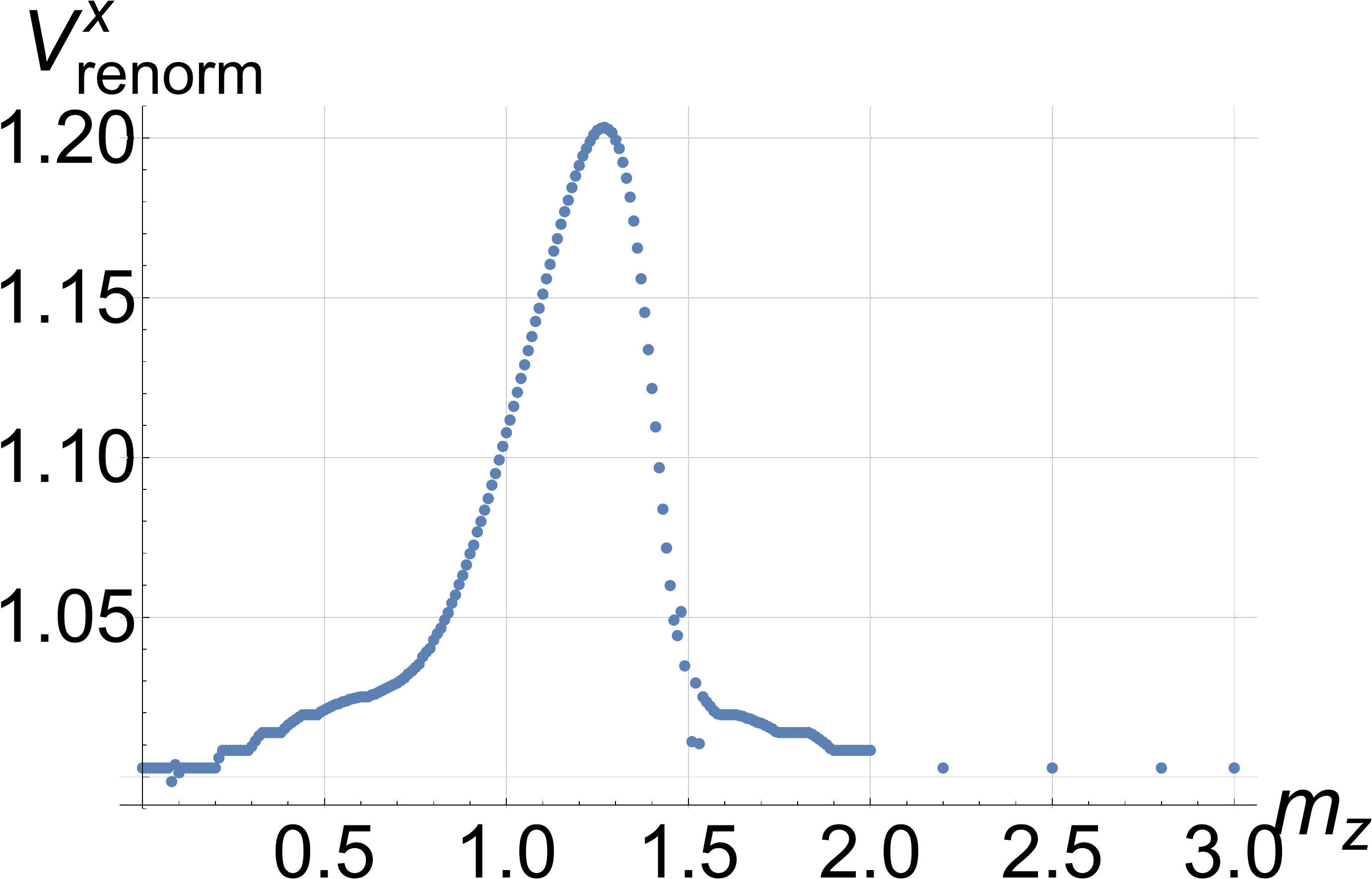}}
  \scalebox{0.155}{\includegraphics{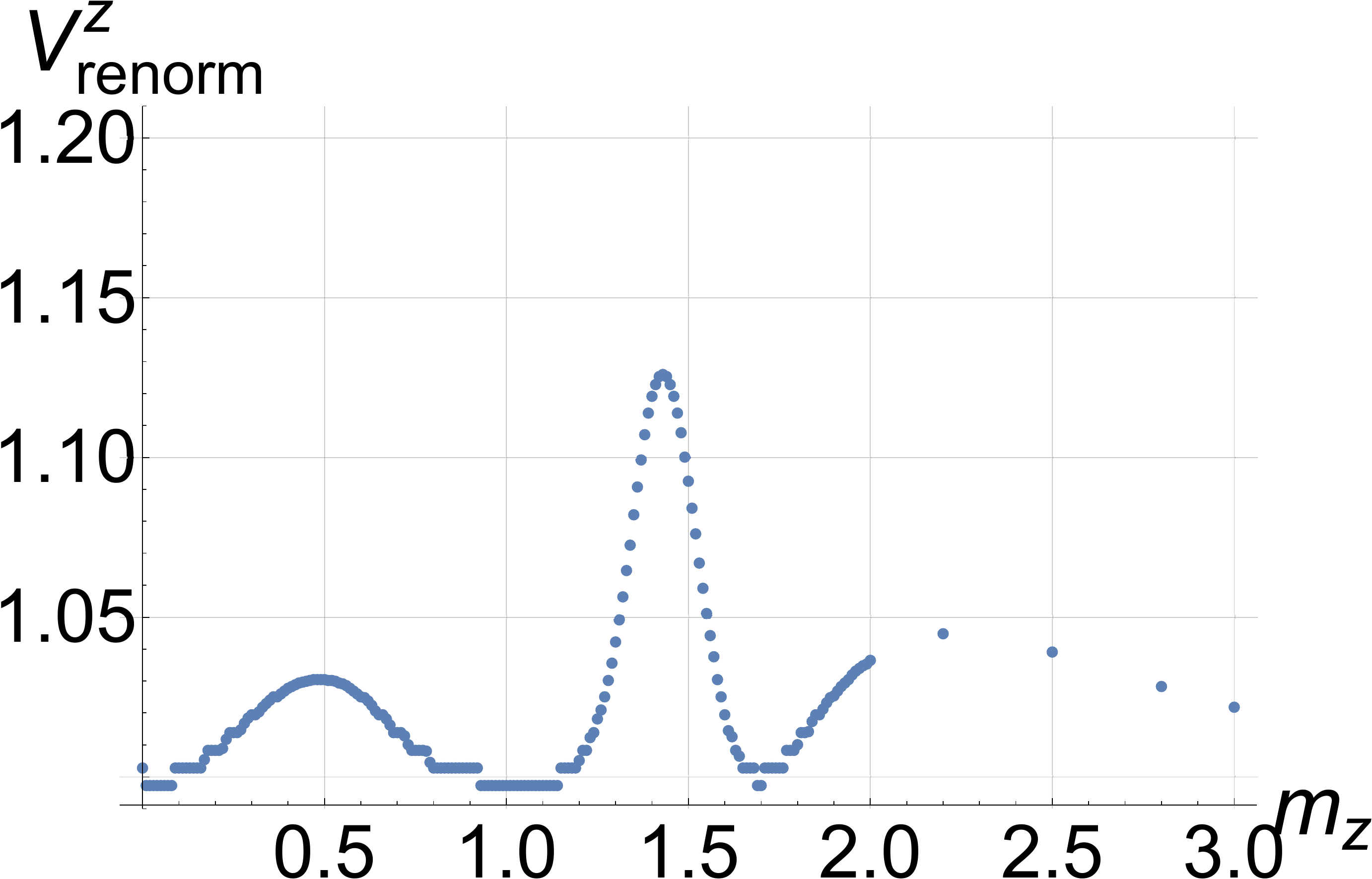}}
\end{minipage}}\\[-10pt]
  \caption{The elastic hardening. $V_\mathrm{renorm}^{x/z}(m_z)$ plots of $R_\mathrm{velo}<1$ and different signs of $\alpha_\frac{\mathrm{DMI}}{\mathrm{EXI}}$. }\label{fig:harden}
\end{figure}

The value of $R_\mathrm{velo}$ strongly affects the excitation spectrum of the effective phonon theory in a clearer manner as compared with $\alpha_{\frac{\mathrm{DMI}}{\mathrm{EXI}}}$. Imagine drawing the original unperturbed phonon dispersion line in Fig.~\ref{fig_magnon}, whose slope might be smaller or larger than that of the blue magnon linear mode, providing a natural bipartite classification. 1) $v_0 < v_\mathrm{mag}$. The phonon mode lies below the blue magnon linear mode and intersects with the green magnon quadratic mode. On one hand, if both the magnon quadratic mode itself and the magnetoelastic interactions are strong enough in intensity, depending on the details of the coupling, the intersection becomes an anticrossing of two reconstructed modes repelling each other in the effective phonon spectrum. On the other hand, near the structural phase transition where fluctuations become immense, the blue magnon mode strongly repels the phonon mode downward as a result of the magnon-phonon interaction, serving as the major cause of the softening effect in Fig.~\ref{fig:soften} in agreement with the experimental data shown Fig.~\ref{fig:softenExpe}. 2) $v_0 > v_\mathrm{mag}$. The phonon mode lies above the blue magnon linear mode and intersects with the orange magnon gapped mode. 
In spite of this, the high energy scale of the gap renders itself irrelevant for the long wavelength phonons. Similar to 1), near the phase transition, the blue mode repels the phonon mode upward, which is an elastic hardening prediction from our theory. As seen in Fig.~\ref{fig:harden}, the renormalized phonon velocity is in general larger than its original unity value. It is expected to be realized by changing either $v_0$ or $v_\mathrm{mag}$. Recent studies of controllable DMI in Skyrmion lattices by exerting strains\cite{anticrossing1,strain_control} or varying compositions\cite{size_control} can be candidates for realizing this case. Last but not least, when $v_0 \approx v_\mathrm{mag}$, the foregoing distinction becomes vague. As a minor reassurance, we indeed saw evident, albeit complex transitions from softening to hardening in $V_\mathrm{renorm}(m_z)$ when $R_\mathrm{velo}$ traverses the range $[0.7,1.3]$.

We discuss more details about the comparison between theory and experiment. Figure \ref{fig:soften} shows the theoretical result that reproduces the experimentally observed signals in Fig.~\ref{fig:softenExpe}, including the drastic decrease in the velocity of the $k_\parallel$-mode and the multi-peak-and-trench fine structure for the $k_\perp$-mode. In this model calculation, we cannot produce a temperature dependence simply because the spectral function in this case doesn't depend on temperature. However, as suggested by basically the same feature at different temperatures marked by the gray dots in Fig.~\ref{fig:softenExpe}, the current theory is able to illustrate the essential physics therein. The external magnetic field in Fig.~\ref{fig:softenExpe} is in general not simply proportional to the parameter $m_z$ in Fig.~\ref{fig:soften} since $m_z$ should be regarded as an approximation in modeling the effect on the deformation of the magnetic structure due to the possibly complex magnetization process. Nevertheless, this comparison suffices to highlight the key magnetoelastic responses.
The reliability of this result is supported by the fact that within the range $\alpha_\frac{\mathrm{DMI}}{\mathrm{EXI}} \in [0.8,1.2] \,, R_\mathrm{velo} \in [1.6,4]$ of the control parameters, the basic characters hold all along. 
After all, we can notice that in Fig.~\ref{fig:soften} the most evident changes in the velocity (stiffness) always reside around the destruction of the monopole lattice, i.e., the monopole-antimonopole pair annihilation at $m_z=\sqrt{2}$\cite{Resistivity}, and especially this topological phase transition clearly manifests itself by the drastic softening in Fig.~\ref{fig:softenzz}. As for the magnitude of the softening with respect to the original $v_0=1$ situation, our theory produces $\frac{\Delta V_\mathrm{renorm}^z}{\Delta V_\mathrm{renorm}^x} \lesssim 5$, which is a bit smaller than the experimental value between 6 and 10. Despite this discrepancy, we highlight the excellent consistency between the theoretical and experimental fine structures. Besides the clear match for single-trench $k_\parallel$-mode case of Fig.~\ref{fig:softenzz}, all three trenches and two peaks in Fig.~\ref{fig:softenxx} find their counterparts in the experiment. 
On the other hand, 
if we set $\alpha_\mathrm{EXI}$ or $\alpha_\mathrm{DMI}$ to zero, it becomes impossible to reproduce the experimental signals. 
Taken as a whole, these strongly endorse our theory and can be regarded as a new way to find out some quantities temporarily outside the reach of experimental detection.

Another aspect worth discussing is why the experimental features are reproduced when $\alpha_\frac{\mathrm{DMI}}{\mathrm{EXI}}$ is not far away from unity although we have $J=10D$. The resolution consists in DMI's unique sensitivity to minute strains\cite{anticrossing1,anticrossing2,strain_control}. Some anticrossing points mixing up spin-up and spin-down bands in the band structure contribute massively to the spin-orbit interaction hence the DMI. The phonon induced strain modifies the band structure slightly, which, however, may considerably change the DMI because the position of the Fermi level in regard to the nearby anticrossing points can drastically change the contribution from these points. Experimentally, typically ten times larger relative modulation of DMI than that of EXI is observed\cite{strain_control}, which is consistent with our theoretical finding.

\subsection{Magnetic-field-dependent evolution of the hybridized excitation spectra}\label{Sec:spectra}
\begin{figure*}
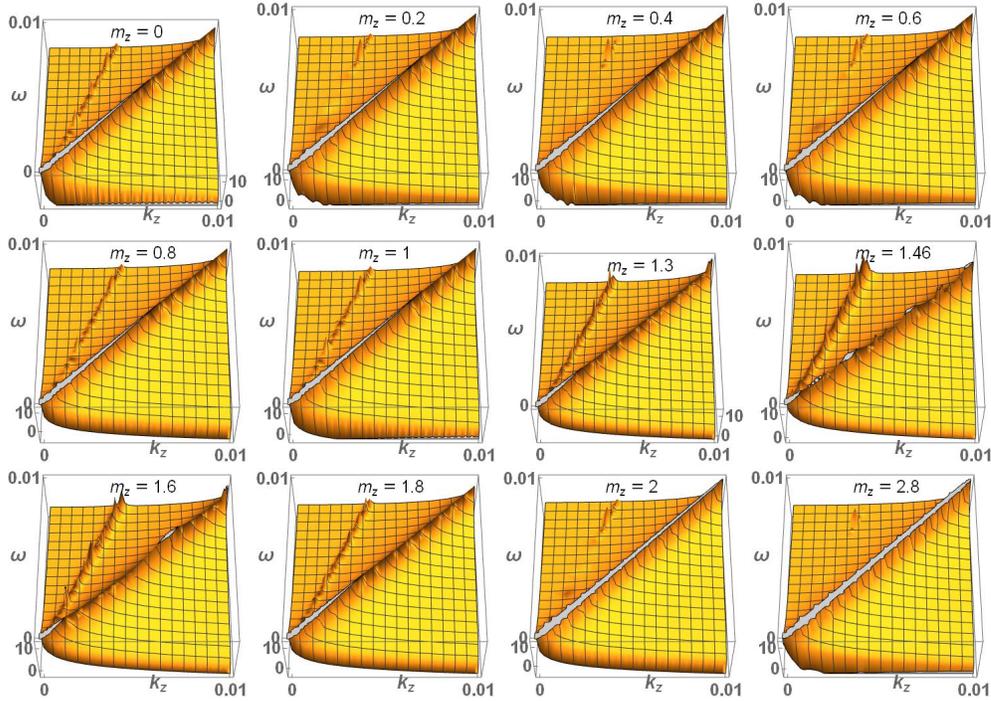

  \scalebox{0.25}{\includegraphics{{{spectralzz_mz0}}}}
  \scalebox{0.25}{\includegraphics{{{spectralzz_mz0.2}}}}
  \scalebox{0.25}{\includegraphics{{{spectralzz_mz0.4}}}}
  \scalebox{0.25}{\includegraphics{{{spectralzz_mz0.6}}}}
  \scalebox{0.25}{\includegraphics{{{spectralzz_mz0.8}}}}
  \scalebox{0.25}{\includegraphics{{{spectralzz_mz1}}}}
  \scalebox{0.25}{\includegraphics{{{spectralzz_mz1.3}}}}
  \scalebox{0.25}{\includegraphics{{{spectralzz_mz1.46}}}}
  \scalebox{0.25}{\includegraphics{{{spectralzz_mz1.6}}}}
  \scalebox{0.25}{\includegraphics{{{spectralzz_mz1.8}}}}
  \scalebox{0.25}{\includegraphics{{{spectralzz_mz2}}}}
  \scalebox{0.25}{\includegraphics{{{spectralzz_mz2.8}}}}
  \caption{Logarithmic plots of spectral function $A_\mathrm{zz}(k_z,\omega)$ at various uniform magnetization $m_z$'s for the $k_\parallel$-mode sound wave.}\label{fig_spectralzz}
\end{figure*}

\begin{figure}
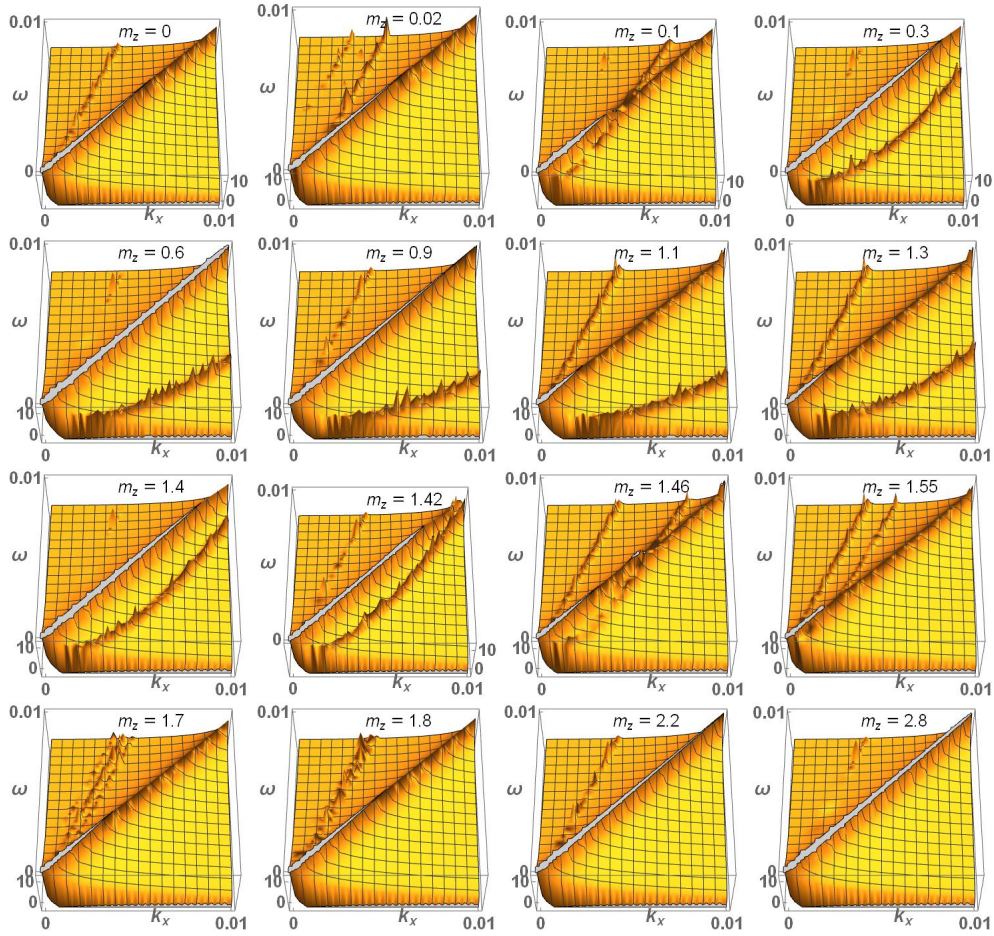

  \scalebox{0.25}{\includegraphics{{{spectralxx_mz0}}}}
  \scalebox{0.25}{\includegraphics{{{spectralxx_mz0.02}}}}
  \scalebox{0.25}{\includegraphics{{{spectralxx_mz0.1}}}}
  \scalebox{0.25}{\includegraphics{{{spectralxx_mz0.3}}}}
  \scalebox{0.25}{\includegraphics{{{spectralxx_mz0.6}}}}
  \scalebox{0.25}{\includegraphics{{{spectralxx_mz0.9}}}}
  \scalebox{0.25}{\includegraphics{{{spectralxx_mz1.1}}}}
  \scalebox{0.25}{\includegraphics{{{spectralxx_mz1.3}}}}
  \scalebox{0.25}{\includegraphics{{{spectralxx_mz1.4}}}}
  \scalebox{0.25}{\includegraphics{{{spectralxx_mz1.42}}}}
  \scalebox{0.25}{\includegraphics{{{spectralxx_mz1.46}}}}
  \scalebox{0.25}{\includegraphics{{{spectralxx_mz1.55}}}}
  \scalebox{0.25}{\includegraphics{{{spectralxx_mz1.7}}}}
  \scalebox{0.25}{\includegraphics{{{spectralxx_mz1.8}}}}
  \scalebox{0.25}{\includegraphics{{{spectralxx_mz2.2}}}}
  \scalebox{0.25}{\includegraphics{{{spectralxx_mz2.8}}}}
  \caption{Logarithmic plots of spectral function $A_\mathrm{xx}(k_x,\omega)$ at various uniform magnetization $m_z$'s for the $k_\perp$-mode sound wave.}\label{fig_spectralxx}
\end{figure}

Obviously, noticing the range of $R_\mathrm{velo}$, Fig.~\ref{fig:soften} belongs to the foregoing $v_0 < v_\mathrm{mag}$ case. A careful inspection of the dispersion profiles, i.e., spectral function plots Fig.~\ref{fig_spectralzz} ($k_\parallel$-mode) and Fig.~\ref{fig_spectralxx} ($k_\perp$-mode), leads us to the following explanation.
For the $k_\parallel$-mode, the dominant modes appear to be the linear excitations of magnons (upper branch) and phonons (lower branch). Increasing $m_z$, the slope of the magnon mode almost remains the same while its intensity gradually increases ($m_z = 0.6\textrm{--}1.46$) to some maximum value around the phase transition and drops down afterward ($m_z = 1.6\textrm{--}2.0$). The larger the intensity becomes, the stronger repulsion is exerted to the phonon mode underneath, explaining the drastic softening. 
The reason why this case lacks the participation of the magnon quadratic mode lies in the magnetoelastic interaction in Eq.~\eref{eq-ME}. The magnon quadratic mode originates from the $\phi$-quadratic term in our spin-wave theory in Eq.~\eref{eq_Lspinwave}. Thus, if it were to largely affect the phonon linear mode in the effective phonon theory, $\phi$-field must have an adequate coupling with phonons. The relevant form factors hereof, $C_\mathrm{EXI/DMI}^{\mu j}$ when $\mu=1,2,3$, is diagonal as seen from the calculation, making the $k_\parallel$-mode ($k_\perp$-mode) phonon primarily couples with $\phi_z$ ($\phi_x$). But only $\phi_x$ and $\phi_y$ in the $\phi$-quadratic term are important since $b_x$ and $b_y$ are vanishing.

For the $k_\perp$-mode, the interplay takes place mainly between the original phonon linear mode and the magnon linear and quadratic modes as seen from their high intensities in Fig.~\ref{fig_spectralxx}. We can first faintly recognize the three magnon modes (gapped, linear and quadratic, from left to right) together with the always dominant phonon linear mode when $m_z = 0.02$, which is in contrast to the $m_z = 0$ case with only one magnon linear dispersion present. This is because of the degeneracy of magnon excitations when $m_z = 0$ mentioned in Sec.~\ref{new_spectrum}. Note that the gapped magnon mode disappears in most plots since increasing $m_z$ enlarges the gap beyond the plot range. For $m_z=0.02$ and $m_z=0.1$, a typical (anti)crossing or reconstruction of the magnon and phonon linear modes occurs at some low-energy scale inside the plot range. The mutual repulsion between them makes the reconstructed phonon linear mode, especially the part right to the crossing point, move downward, explaining the first small trench in Fig.~\ref{fig:softenxx}. Once the crossing point gets higher ($m_z=0.3$), the low-energy part of the phonon mode simply bounces back and gives rise to the first small peak near $m_z=0.25$ in Fig.~\ref{fig:softenxx}. Thereafter, as $m_z$ increases till the vicinity of $1.2$, the magnon linear mode above the phonon linear mode becomes stronger and stronger while the magnon quadratic mode keeps moving down till $m_z=1.1$, which constructively pushes the phonon mode downward, yielding the deepest trench. And then, abruptly, a transient reverse procedure is observed approximately from $m_z=1.1$ to $m_z=1.42$ plots, hence the dramatic upsurge in Fig.~\ref{fig:softenxx}. Next, the magnon quadratic mode recovers all the way back and crosses over the phonon mode at lower and lower energy scales while the magnon linear mode becomes stronger till $m_z=1.55$, which is again a constructive effect of dragging downward the phonon mode. Finally, a fading reunion of the three magnon modes are observed from $m_z=1.7$ to $m_z=2.2$, which is natural for an induced ferromagnetism of too large $m_z$.


In a nutshell, the experimentally observed magnetoelastic phenomena are the consequences of two aspects that vary with $m_z$ or the external magnetic field. One is the magnon quadratic mode generated at nonzero $m_z$ (together with the herein insignificant gapped mode), whose reciprocating shift in the spectral function plot in Fig.~\ref{fig_spectralxx} is clearly controlled by the emergent magnetic field $b_z(m_z)$ as explained in Sec.~\ref{new_spectrum} in terms of Eq.~\eref{eq_Lspinwave}. The other is the family of form factors directly affecting the intensities of and hybridization between various modes. They show nonmonotonous behavior upon increasing $m_z$ and vary strongly near the phase transition. 
An intricate integration of these two aspects leads to the rich experimental features. A detailed inspection of the underlying ground state spin configuration provides more insights\cite{Resistivity}. Here we only recapitulate two key aspects. One is the nonmonotonic profile of $b_z(m_z)$, whose maximum near $m_z=1.0$ and fast dip around $m_z = \sqrt{2}$ make the form factors' variation more perceivable. Note that the realistic lattice cutoff introduced to the monopole defects in the calculation will also postpone the complete destruction of the monopole lattice to some value larger than the ideal value $m_z=\sqrt{2}$. The other, from a more intriguing viewpoint, is the nontrivial contribution from the (anti)monopoles and their characteristic collision-and-annihilation motion. In contrast to the SkX in MnSi discussed below, we have argued for the crucial role of the topological defects, i.e, the (anti)monopoles, in the magnetoresistivity and especially the topological phase transition of the destruction of the monopole lattice. Here in the magnetoelastic phenomena, not only $b_z$ but also the form factors are reflecting the rich facets of the spin textures. It is the coupling between phonons and the (anti)monopoles that causes all the complexities.


\subsection{Relation to triangular lattice of Skyrmion tubes in MnSi}\label{result_MnSi}
Now we briefly discuss the magnetoelastic couplings in the SkX of MnSi. 
Except from the modification in the spin-wave theory, we can apply the similar form of couplings to MnSi. 
Nevertheless, we observe vanishing couplings when the sound wave propagates along the cylindrical symmetry $z$ axis of Skyrmion tubes in MnSi. For the EXI induced coupling, the form factors are simply zero due to the translational symmetry along the $z$ direction. Although this is not the case for the coupling induced by DMI, the spatial averages of corresponding form factors turn out to be 
zero in the end. As for the perpendicular propagating case, we observe nonvanishing magnetoelastic effects, albeit negligibly smaller as compared to MnGe. 
On the other hand, the experimental signals ($\frac{\Delta \kappa}{\kappa}\approx 0.1\%$) of the SkX in MnSi are quite smaller than those ($\frac{\Delta \kappa}{\kappa} \approx 2\% \sim 10\%$) detected in MnGe\cite{Nii2}. 
Therefore, our form of magnetoelastic couplings turns out to be the leading order effect in the MnGe monopole lattice. Based on the solid state mechanics and macroscopic thermodynamics, many more complex higher order terms and fitting parameters are incorporated into a Ginzburg-Landau free energy calculation\cite{MnSi_Ela_theory} for the ultrasonic signals in MnSi\cite{Nii1,Petrova}. 
This comparison actually lends credence to the aforementioned essential role played by the monopole defects that are absent in MnSi. In other words, the contribution from monopole defects to the magnetoelastic effects, if present at all, dominates and is captured by the formalism developed in this work.

\section{Concluding remark}\label{conclusion}
The study of ultrasonic elastic responses possesses strong motivation from the experimental findings. We not only explain the observed softening effect but also predict new issues, e.g., hardening and other patterns of the dependence of the stiffness on the magnetic field, which in return provides a way to determine some experimentally inaccessible physical quantities. Based on a well-established spin-wave theory from the previous magnetoresistivity study, we are able to identify once again the nontrivial features of the monopole lattice, especially the topological phase transition signifying strong correlations. Thanks to the agreement with the experimental observations, this magnetoelasticity study, together with the magnetoresistivity one, can establish the importance of the topological nature of the spin configuration in strongly correlated electronic systems. As a whole, they speak for a crucial role played by the monopole defects in chiral magnet MnGe. In particular, these studies pave the way for even more intriguing scenarios of coupling topologically nontrivial objects with other systems. They show the rich physics therein and help us gain insights for further investigations towards plenty of manipulation methods for Skyrmionics applications.

\section*{Acknowledgments}
We thank Yoichi Nii for useful discussions and the indispensable experimental results. X.-X.Z was partially supported by the Panasonic Scholarship and by Japan Society for the Promotion of Science through Program for Leading Graduate Schools (ALPS) and Grant-in-Aid for JSPS Fellows (No. 16J07545). This work was supported by JSPS Grant-in-Aid for Scientific Research (No. 24224009) and JSPS Grant-in-Aid for Scientific Research on Innovative Areas (No. 26103006) from MEXT, Japan, and ImPACT Program of Council for Science, Technology and Innovation (Cabinet office, Government of Japan). This  work  was  also  supported  by  CREST, Japan Science and Technology Agency.

\label{Bibliography}
\bibliographystyle{unsrt}
\bibliography{reference.bib}  

\end{document}